\documentclass[12pt]{article}

\usepackage{amsfonts,amssymb,amsmath}
\usepackage[dvips]{epsfig}
\begin{document}
\title{\bf Effects of oscillatory deformations on the coherent and incoherent quantum transport}
\author{ Naghi Behzadi $^{a}$
\thanks{E-mail:n.behzadi@tabrizu.ac.ir} and
Bahram Ahansaz $^{b}$
\thanks{E-mail:b.ahansaz@azaruniv.edu}
\\ $^a${\small Research Institute for Fundamental Sciences, University of Tabriz, Tabriz, Iran,}
\\ $^b${\small Department of Physics, Shahid Madani University of Azarbayijan, Tabriz, Iran.}} \maketitle

\begin{abstract}
\noindent
Inspired by the works of [F. Caruso, New J. Phys. 16, 055015 (2014)] and [T. Scholak $et$ $al$, J. Phys. B: At. Mol. Opt. Phys. 44 184012 (2011)], which state that for a large class of complex noisy networks, the optimal efficiency of quantum transport is universally obtained by mixing coherent (Hamiltonian) and incoherent (noisy) parts where the contribution of the coherent part is strictly more than incoherent one, we examine the effect of oscillatory deformations on two simple prototypes in order to study their effects on the efficiency of coherent and incoherent energy transport. The prototypes are interchangeable to each other only by a simple phase modulation, such that the dynamics for the first type is only coherent, while for the second one the coherent evolution is completely suppressed and the evolution of the system is only incoherent (noisy). In this regard, it is shown that there exist a special deformation by which the efficiency of incoherent transport becomes better than the coherent one. This result suggests that in the noisy networks with collective harmonic motions, the optimality of transport can be occurred in such a way that the contribution of incoherent term is more than the coherent one.
\\
\\
{\bf PACS Nos:}
\end{abstract}

\section{Introduction}
Transport phenomena have been central to quantum mechanics since its early days.
Recently, it has been renewed by the prospect of transferring quantum information across quantum networks \cite{Hartley, kay, vah, beh} and
the recurring interest in understanding the fundamental processes, which influence
energy transport in photosynthetic systems \cite{moh, Sension, Plenio1, Plenio2, Plenio3}. The presence of environmental noises is generally considered to be an unavoidable hindrance
for efficient transport of charge or energy through quantum systems and the general view is
that transport in quantum systems relies on their coherence, which is inevitably reduced by
interactions with an external noisy environment. However, inspired by the experimental results, further theoretical studies on energy transport in light
harvesting complexes have been carried out to investigate the role of noises, and particularly the dephasing ones,
in the process of exciton transport in these complexes \cite{reb, kassal, Sin, lim}.
Indeed, the efficient transport observed in certain biological systems is not compatible with a fully
coherent evolution, so in this way, the interplay between coherent (unitary) dynamics and incoherent (irreversible) dynamics gives the optimal way for quantum transport in many noisy systems.
Recently, it has begun to be appreciated that vibrational modes arising in molecular structures may play an important role in the dynamics of such systems \cite{rei, iri}. What is clear particularly at room temperature is that, exciton energy transport depends not only
on the topology of electronic couplings among sites, but also on the simultaneous effects of the molecular
motions and environmental fluctuations, which drive efficient
transport processes. Also, collective vibrational motions which may arise through a coupled many-body quantum system can lead to an enhancement in the transport of excitations across such systems \cite{sem, asa}.

On the other hand, the optimal mixing of coherent and incoherent quantum transport depends on the initial state preparation $\cite{Caruso, cui}$. In Ref. $\cite{Caruso}$, it was shown that, when the excitation is initially prepared in one end of a linear chain, the optimal transport of excitation through it is only coherent. In other words, optimal transport is only obtained by  self evolution of the system. Also, in Ref. $\cite{beh2}$, incoherent quantum transport in regular networks has been investigated, where the coherent transport is completely suppressed due to destructive interferences. In this case, the optimality of the incoherent quantum transport depends on the optimal effects of dephasing noises.

In the present work, we consider two simple networks, each containing four two-level systems and are interchangeable to each other only by a phase modulation on one of the coupling strength. Each network has an additional dissipative sink site attached to it. For the first network, the optimal dynamics is only coherent, while for the second one is only incoherent, which in turn, is related to the optimal effects of dephasing noises on the system. In this situation, we consider some harmonic oscillatory deformations on the geometry of configurations and highlight their effects on the respective efficiency of transports. It is shown that in the absence of these deformations, the efficiency of coherent transport is better than the incoherent one, which is in accordance with the results of Ref. $\cite{Caruso}$. However, in the presence of harmonic deformations, it is observed that the efficiency of incoherent transport can be improved to be better than the coherent one. These results, in turn, ensure the point that the induced evolutions can be more effective than the self evolutions of a system; a fact which can be observed in biological systems.

This paper is organized as follows: In Sec. 2, we demonstrate the basic ingredients for achieving the optimal coherent and incoherent quantum transport through two types of configurations with fixed vertices, along with making a comparison between them. Sec. 3 is devoted to describe the various useful harmonic deformations, which can occur in the structure of configurations and to explain their effects on the respective efficiency of coherent and incoherent quantum transport. Finally, a brief conclusion is presented in Sec. 4.

\section{The model}
The general Hamiltonian describing the energy transport of an excitation through a network composed of four two-level quantum system, as depicted in Fig. 1(a) and Fig. 2(a), is given as follows
\begin{eqnarray}
&&\hspace{-7mm}{H}=\sum_{i=1}^{4} \hbar\omega_{i}{\sigma}_{i}^{+}{\sigma}_{i}^{-} +\sum_{\{i, j\}\in E}\hbar J_{i,j} \left({\sigma}_{i}^{-}{\sigma}_{j}^{+}+{\sigma}_{i}^{+}{\sigma}_{j}^{-}\right),
\end{eqnarray}
where $\sigma^{+}_{i}=|i\rangle\langle0|$ and $\sigma^{-}_{i}=|0\rangle\langle i|$
are the raising and lowering operators for a two-level system lied at $i$th vertex of the network with transition frequency $\omega_{i}$. We assume that the atoms are identical and so we have $\omega_{1}=\omega_{2}=\omega_{3}=\omega_{4}=\omega$. The strength of coupling between the $i$th and $j$th atoms is denoted by $J_{i,j}$, which indicates the hopping rate of excitation between them. $E$ is the set of network edges, corresponding to the coupling between the sites, as shown in Fig. 1(a) and Fig. 2(a).
We consider a configuration in which, all of the coupling constants are equal to each other, i.e.
\begin{eqnarray}
J_{1,2}=J_{1,3}=J_{2,4}=J_{3,4}=J,
\end{eqnarray}
(see Fig. 1(a)). This corresponds to the network in which, the optimal dynamics is coherent. To clarify this point, let us introduce a new set of basis as \cite{chris}:
\begin{eqnarray}
\begin{array}{c}
  |s_{1}\rangle:=|1\rangle, \quad |s_{2}\rangle:=\frac{1}{\sqrt{2}}(|2\rangle+|3\rangle), \quad |s_{3}\rangle:=|4\rangle,
\end{array}
\end{eqnarray}
where $|s_{1}\rangle$, $|s_{2}\rangle$ and $|s_{3}\rangle$ are set of basis corresponding to the three column of Fig. 1(a) (see Ref. \cite{chris}). The Hamiltonian (1) in this basis becomes as
\begin{eqnarray}
&&\hspace{-7mm}{H}=\sqrt{2}J(|s_{1}\rangle\langle s_{2}|+|s_{2}\rangle\langle s_{3}|)+h.c..
\end{eqnarray}
Eq. (4) is clearly similar to the Hamiltonian of a three site chain with equal coupling $\sqrt{2}J$ (see Fig. 1(b)). As denoted in Ref. $\cite{Caruso}$, if the excitation for this configuration is initially prepared at site 1, the coherent evolution is the optimal dynamics in transferring it to the sink or reaction center.

Now let us consider the phase modulation $J_{3, 4}\rightarrow-J_{3, 4}$ in (2), which corresponds to the other our demanding network in this paper (see Fig. 2(a)). For this case, we introduce another set of basis in the single excitation subspace, \cite{kay, vah, beh, beh2}, as
\begin{eqnarray}
\begin{array}{c}
  |s_{1}\rangle:=|1\rangle, \quad |s^{\pm}_{1}\rangle:=\frac{1}{\sqrt{2}}(|2\rangle\pm|3\rangle), \quad |s_{2}\rangle:=|4\rangle.
\end{array}
\end{eqnarray}
The Hamiltonian (1), indeed, is left with a direct sum structure as
\begin{eqnarray}
H=H_{1}\bigoplus H_{2},
\end{eqnarray}
where
\begin{eqnarray}
\begin{array}{c}
  H_{1}=\sqrt{2}J(|s_{1}\rangle\langle s^{+}_{1}|+|s^{+}_{1}\rangle\langle s_{1}|),\\\\
  H_{2}=\sqrt{2}J(|s^{-}_{1}\rangle\langle s_{2}|+|s_{2}\rangle\langle s^{-}_{1}|),
\end{array}
\end{eqnarray}
and the respective invariant subspaces are denoted as below
\begin{eqnarray}
\begin{array}{c}
  \mathcal{H}_{1}=\mathrm{span}\{|s_{1}\rangle, |s^{+}_{1}\rangle\},\\\\
  \mathcal{H}_{2}=\mathrm{span}\{|s^{-}_{1}\rangle, |s_{2}\rangle\}.
\end{array}
\end{eqnarray}
It is well-known that, for this network, coherent evolution can not transfer the excitation from site 1 to the reaction center (see Fig. 2(b)). As the coherent evolution of the system is only restricted to the invariant subspace $\mathcal{H}_{1}$, therefore, the existence of induced or incoherent evolution arisen from the system interaction with fluctuating environments, which conserves the energy, is necessary. We consider, without loss of generality, that the second network interacts with the structureless environments through the sites 2 and 3. Therefore, in the Markovian approximation, the effects of these interactions on the dynamics of this system, called dephasing noises, are described by the following Lindblad super-operator
\begin{eqnarray}
  \mathcal{L}_{deph}(\rho)=\sum_{i=2}^{3} \gamma_{i}(2{\sigma}_{i}^{+}{\sigma}_{i}^{-}\rho{\sigma}_{i}^{+}{\sigma}_{i}^{-}-\{{\sigma}_{i}^{+}{\sigma}_{i}^{-},\rho\}),
\end{eqnarray}
where $\gamma_{i}$s are the rates of dephasing noises, which randomize the phases of local excitations
and $\{A,B\}:=AB+BA$.

In order to measure how much of the excitation energy is transferred along the both networks, we introduce an additional site, the sink, which is connected
to site $4$. The sink is populated by an irreversible decay process from a
chosen site, as described by
\begin{eqnarray}
&&\hspace{-7mm}\mathcal{L}_{sink}(\rho)=\Gamma(2{\sigma}_{5}^{+}{\sigma}_{4}^{-}\rho{\sigma}_{4}^{+}{\sigma}_{5}^{-}
-\{{\sigma}_{4}^{+}{\sigma}_{5}^{-}{\sigma}_{5}^{+}{\sigma}_{4}^{-},\rho\}),
\end{eqnarray}
where $\Gamma$ is the rate of dissipative irreversible process, that reduces the number of excitations in the system.
Therefore, the population of the sink, referred to as transport efficiency, is given by
\begin{eqnarray}
P_{sink}(t)=2 \Gamma\int_{0}^{t}\rho_{_{4,4}}(t')dt'.
\end{eqnarray}
We note, in this paper, that the Lindblad operators are time-independent and the dephasing rates are positive, i.e. $\gamma_{i}\geq0$, therefore the dynamics for the second network, as an open system, can be described by time-independent Markovian master equation in the following Lindblad form $\cite{liane}$
\begin{eqnarray}
\frac{d\rho}{dt}=-i[H ,\rho]+\mathfrak{L}(\rho),
\end{eqnarray}
where $\mathfrak{L}(\rho)=\mathcal{L}_{deph}(\rho)+\mathcal{L}_{sink}(\rho)$.
Remember that, we exploit $\mathfrak{L}(\rho)$ in the master equation for the first network without the term $\mathcal{L}_{deph}(\rho)$.

When all of the sites are fixed, both systems are initially prepared with a single excitation localized
at the site 1, i.e. $\rho(0)=|1\rangle \langle1|$. For the second system, it is assumed that $\gamma_{2}=\gamma_{3}=\gamma$ and $\Gamma=2\gamma$ and for it, the maximal efficiency of incoherent transport occurs for the optimal value of dephasing rate $\gamma=\gamma_{\mathrm{opt}}=1.05$. As a further illustration, it should be noted that for the second network, existence of incoherent transport strongly depends on the dephasing noises arisen from coupling of the system with independent fluctuating environments through the sites 2 and 3 with dephasing rates $\gamma_{2}=\gamma_{3}=\gamma$. The dephasing noises randomize the corresponding phases of the non-local excitations as $|s^{+}_{1}\rangle\in\mathcal{H}_{1}$ \cite{beh2}. Therefore, evolution of the excitation from site 1 to the sink is possible by this process. If the rate of dephasing noises increases, the randomization rate of the phases of corresponding excitations in the invariant subspace $\mathcal{H}_{1}$ becomes considerable and so, the efficiency of incoherent transport is improved. On the other hand, with further increasing in $\gamma$, the efficiency drops as noises suppress randomizing process via the quantum Zeno effect \cite{Plenio2, Plenio3}. Consequently, the maximal efficiency of incoherent transport is occurred for a particular value of dephasing rate called as optimal dephasing rate $\gamma_{\mathrm{opt}}$.

Fig. 3 shows the population of the sink versus time for both systems. It is clear that the optimal efficiency of coherent transport is more than the optimal efficiency of incoherent transport. This observations is in complete agreement with results of Refs $\cite{Caruso, scholak}$.

\section{Effects of harmonic oscillatory deformations}

We consider the harmonic oscillatory deformations as mechanical oscillations of the vertices around their respective equilibrium positions, which in turn modulate the distance-dependent dipolar coupling. Time-dependent distance between the sites $i$ and $j$
is as follows
\begin{eqnarray}
&&\hspace{-7mm}d_{i,j}(t)=d_{0}-[u_{i}(t)-u_{j}(t)]=d_{0}(1-2a_{i,j}sin(\omega_{0} t+\phi_{i,j}))
\end{eqnarray}
where $d_{0}$ is the equilibrium distance between two connected sites,
$u_{i}$ is the displacement of the $i$th site from its equilibrium position and $a_{i,j}$ is the individual relative amplitude of oscillations of sites $i$th and $j$th, when they
move with opposite phase around their equilibrium positions.
The dipole-dipole coupling between the sites $i$ and $j$ has the form \cite{asa} of
\begin{eqnarray}
&&\hspace{-7mm}J_{i,j}(t)=\frac{\tilde{J_{0}}}{[d_{i,j}(t)]^{3}}=\frac{J_{0}}{(1-2a_{i,j}sin(\omega_{0} t+\phi_{i,j}))^{3}},
\end{eqnarray}
where $\tilde{J_{0}}$ contains the dipole moments and physical constants. We
define $J_{0}=\tilde{J_{0}}/d_{0}^{3}$, which has the energy unit.
Henceforth, all energies, time scales and rates will be expressed
in units of $J_{0}$. Therefore, by the existence of some time-dependent coupling strength, such as $J_{i,j}(t)$, the equation (1) represents a time-dependent Hamiltonian. Therefore, the deformed time-dependent Hamiltonian for the coherent evolution becomes as

\begin{eqnarray}
&&\hspace{-7mm}{H}^{coh}(t)=\sqrt{2}\zeta_{1}(t)(|s_{1}\rangle\langle s_{2}|+|s_{2}\rangle\langle s_{1}|)+\sqrt{2}\zeta_{2}(t)(|s_{2}\rangle\langle s_{3}|+|s_{3}\rangle\langle s_{2}|),
\end{eqnarray}
and for the incoherent one as
\begin{eqnarray}
H^{incoh}(t)=H_{1}(t)\bigoplus H_{2}(t),
\end{eqnarray}
with
\begin{eqnarray}
\begin{array}{c}
  H_{1}(t)=\sqrt{2}\zeta_{1}(t)(|s_{1}\rangle\langle s^{+}_{1}|+|s^{+}_{1}\rangle\langle s_{1}|),\\\\
  H_{2}(t)=\sqrt{2}\zeta_{2}(t)(|s^{-}_{1}\rangle\langle s_{2}|+|s_{2}\rangle\langle s^{-}_{1}|).
\end{array}
\end{eqnarray}
By these considerations, the time-independent Markovian master equation (12) is replaced by a time-dependent one.

Now let us consider a case in which, only the site 1 oscillates around its equilibrium along the horizontal line connecting the sites 1 and 4, then $\zeta_{1}(t)=J_{1,2}(t)=J_{1,3}(t)$ oscillates harmonically with time, while $\zeta_{2}(t)=J_{2,4}(t)=\pm J_{3,4}(t)=1$ is constant (see Fig. 4(b)). For this case, the optimal transport efficiency for the coherent and incoherent transfer of excitation from site 1 to the sink is shown in Fig. 4 (a). Contrary to the case shown in Fig. 3, we particularly see that as time goes on, the optimal efficiency of incoherent transport becomes better than the coherent one.

As an another observation, consider that only the site 4 oscillates in similar way as of previously mentioned oscillation of the site 1, so that $\zeta_{1}(t)=J_{1,2}(t)=J_{1,3}(t)=1$ and $\zeta_{2}(t)=J_{2,4}(t)=\pm J_{3,4}(t)$ (Fig. 5(b)). Fig. 5(a) shows that the optimal transport efficiency for the coherent and incoherent evolutions is completely different from those depicted in Fig. 4(a). It is, indeed, hard to judge that the efficiency of incoherent transport is more favorable than the coherent one.

The interesting instance that reveals the superiority of incoherent transport on the coherent one occurs when the sites 1 and 4 oscillate simultaneously with phase difference $\Delta\varphi=\pi$ along the horizontal line in such a way that both $\zeta_{1}(t)$ and $\zeta_{2}(t)$ change differently with time, as shown in Fig. 6(b). As time goes on, the transport efficiency for incoherent evolution becomes much better than the coherent one (see Fig. 6(a)). This is an evidence for the preference of the optimal incoherent transport to the coherent one when they are accompanied by this type of harmonic deformations.

The oscillations of site 2 and 3 (or 1 and 4) with the same phase, i.e. $\Delta\varphi=0$, which give the $\zeta_{1}(t)=\zeta_{2}(t)$ (see Fig. 7(b)), do not lead to the improvement of incoherent transport relative to the coherent one, as shown in Fig. 7(a) and consequently, this situation does not provide a new result different from the advantages of $\cite{Caruso, scholak}$.

At the end, it should be remembered that the focus of the paper is on the making a comparison between the pure coherent and incoherent efficiency of quantum transport. So in this regard, it is important to consider deformations in which, the Hamiltonian for the second network is left with a block diagonal form (see Eq. (16)) and the corresponding Hilbert space is decomposed to direct sum of invariant subspaces such as (8). By this condition, the evolution of the excitation from site 1 to the sink is pure incoherent, i.e. it is possible only in the presence of noises imposed to sites 2 and 3. Obviously, if we take deformations other than those considered in this paper, the Hamiltonian for the second configuration can not be reduced to block diagonal form, and similarly the corresponding Hilbert space will be impossible to reduce to direct sum of invariant subspaces. Consequently, the quantum transport becomes a mixture of coherent and incoherent parts instead of being pure incoherent.

\section{Conclusions}
In this paper, we investigated the accompaniment of harmonic oscillatory deformations, which may be created from thermal fluctuations with the optimal coherent and incoherent quantum transport. For the networks discussed in this paper, the optimal transport in the absence of deformations, is only the pure coherent transport. However, we found out that in the presence of some rare harmonic deformations, optimal transport is pure incoherent one. This, indeed, induces the notion that the environmental effects on the quantum transport in many-body quantum systems may be more efficient than the intrinsic quantum mechanical effects in those systems. The other point, which may be interesting in this regard, is the analysis of structural or memory effects of the environments interacting with the second configuration via sites 2 and 3, which can be investigated in future.

\newpage
Fig. 1. (a) Configuration of four interacting two-level atoms with $J_{1,2}=J_{1,3}=J_{2,4}=J_{3,4}=J$, irreversibly connected to the sink site. (b) The equivalent configuration when the set of basis introduced in Eq. 3 is used.
\begin{figure}
\centering
\includegraphics[width=300 pt]{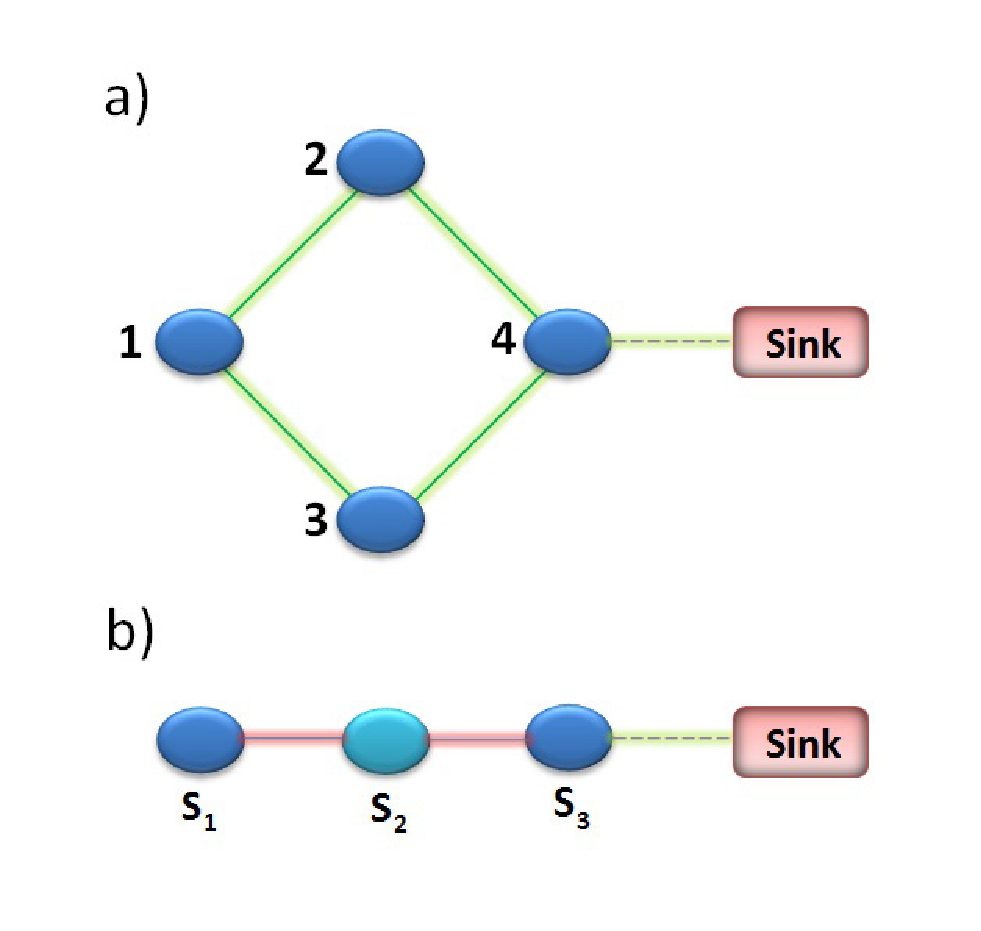}
\caption{} \label{Fig1}
\end{figure}

\newpage
Fig. 2. (a) Configuration of four interacting two-level atoms with $J_{1,2}=J_{1,3}=J_{2,4}=-J_{3,4}=J$, irreversibly connected to the sink site. Dephasing Markovian noises affect the system through site 2 and 3. (b) The equivalent configuration when the set of basis introduced in Eq. (5) is used. Invariant subspaces are connected incoherently to each other by the dephasing noises.
\begin{figure}
\centering
\includegraphics[width=300 pt]{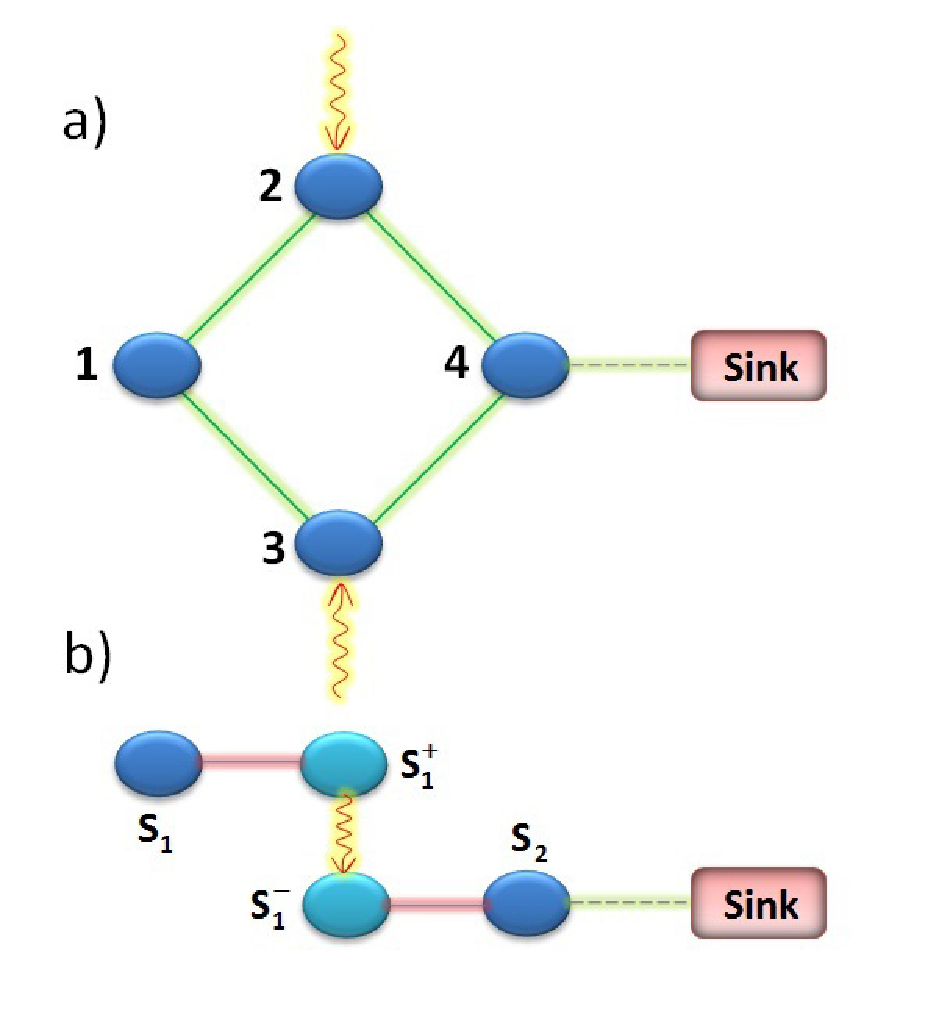}
\caption{} \label{Fig1}
\end{figure}

\newpage
Fig. 3. The solid and dashed blue curves show the populations of the sinks or optimal efficiency of transports for the coherent and incoherent transfer of excitation through first and second configurations represented in Fig. 1 and Fig. 2, respectively, when all of the sites are fixed. The dotted-dashed and the dotted green curves indicate the sum of populations of all sites, for the first and second configurations, respectively.
\begin{figure}
\centering
\includegraphics[width=350 pt]{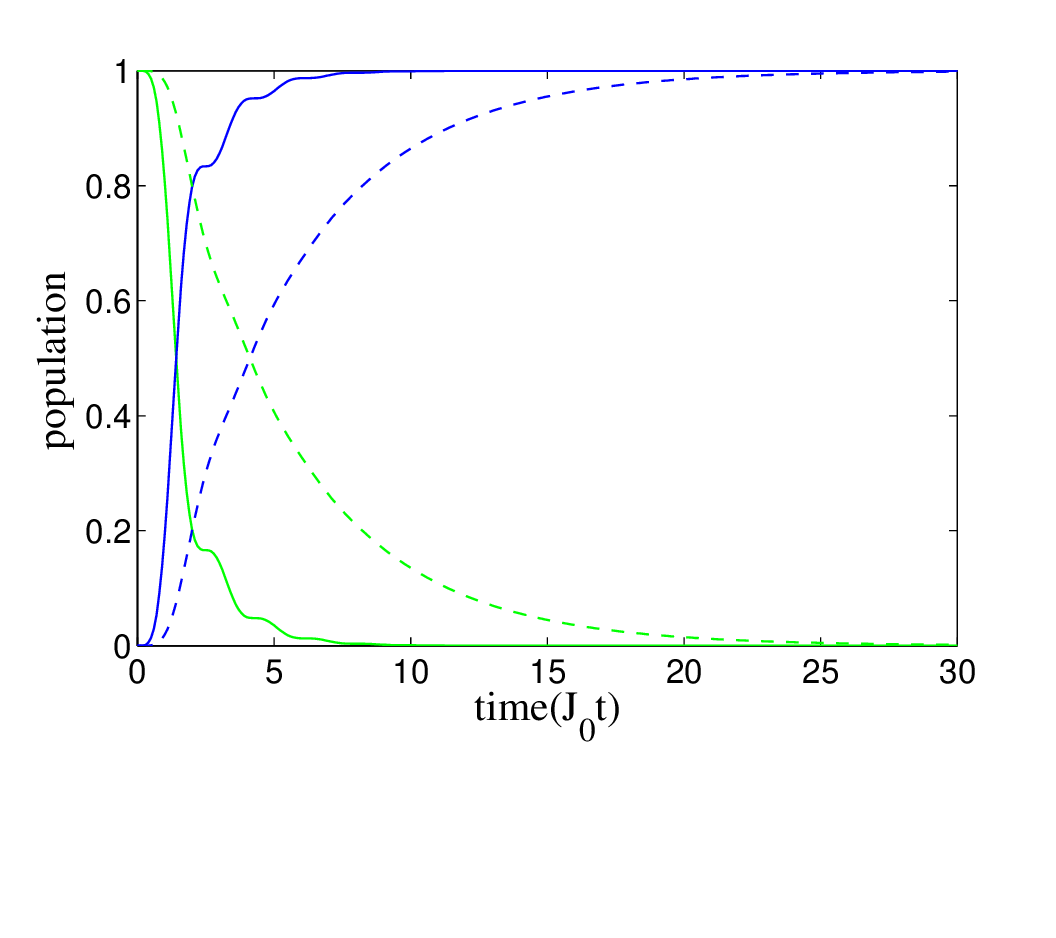}
\caption{} \label{Fig1}
\end{figure}

\newpage
Fig. 4. (a) Populations of the sinks for the optimal coherent (solid blue curve) and optimal incoherent (dashed blue curve) transport, when the site 1 oscillates around its equilibrium along the horizontal line for both configurations with assumptions of $a=1/4$, $\phi=0$ and $\omega_{0}=1$. (b) Time dependence of the coupling strength $\zeta_{1}(t):=J_{1,2}(t)=J_{1,3}(t)$
(solid red curve) and $\zeta_{2}(t):=J_{2,4}(t)=\pm J_{3,4}(t)=1$ (dotted-dashed black curve) where, plus and minus correspond to the coherent and incoherent transports, respectively.
\begin{figure}
\centering
\includegraphics[width=445 pt]{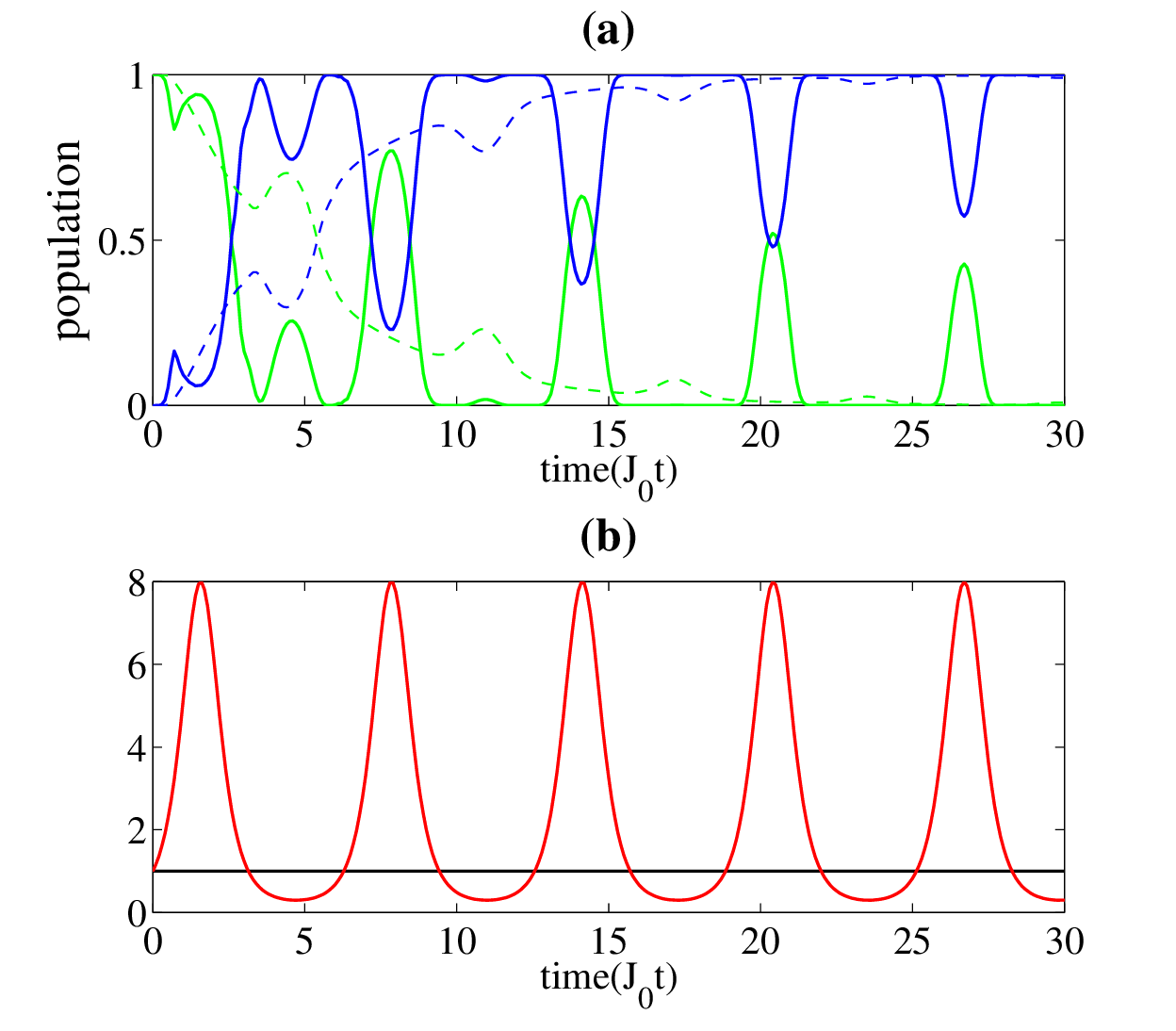}
\caption{} \label{Fig1}
\end{figure}

\newpage
Fig. 5. (a) Populations of the sinks for the optimal coherent (solid blue curve) and optimal incoherent (dashed blue curve) transport, when the site 4 oscillates around its equilibrium along the horizontal line for both configurations with assumptions of $a=1/4$, $\phi=0$ and $\omega_{0}=1$. (b) Time dependence of the coupling strength $\zeta_{1}(t):=J_{1,2}(t)=J_{1,3}(t)=1$ (dotted-dashed black curve) and $\zeta_{2}(t):=J_{2,4}(t)=\pm J_{3,4}(t)$ (solid red curve).
\begin{figure}
\centering
\includegraphics[width=445 pt]{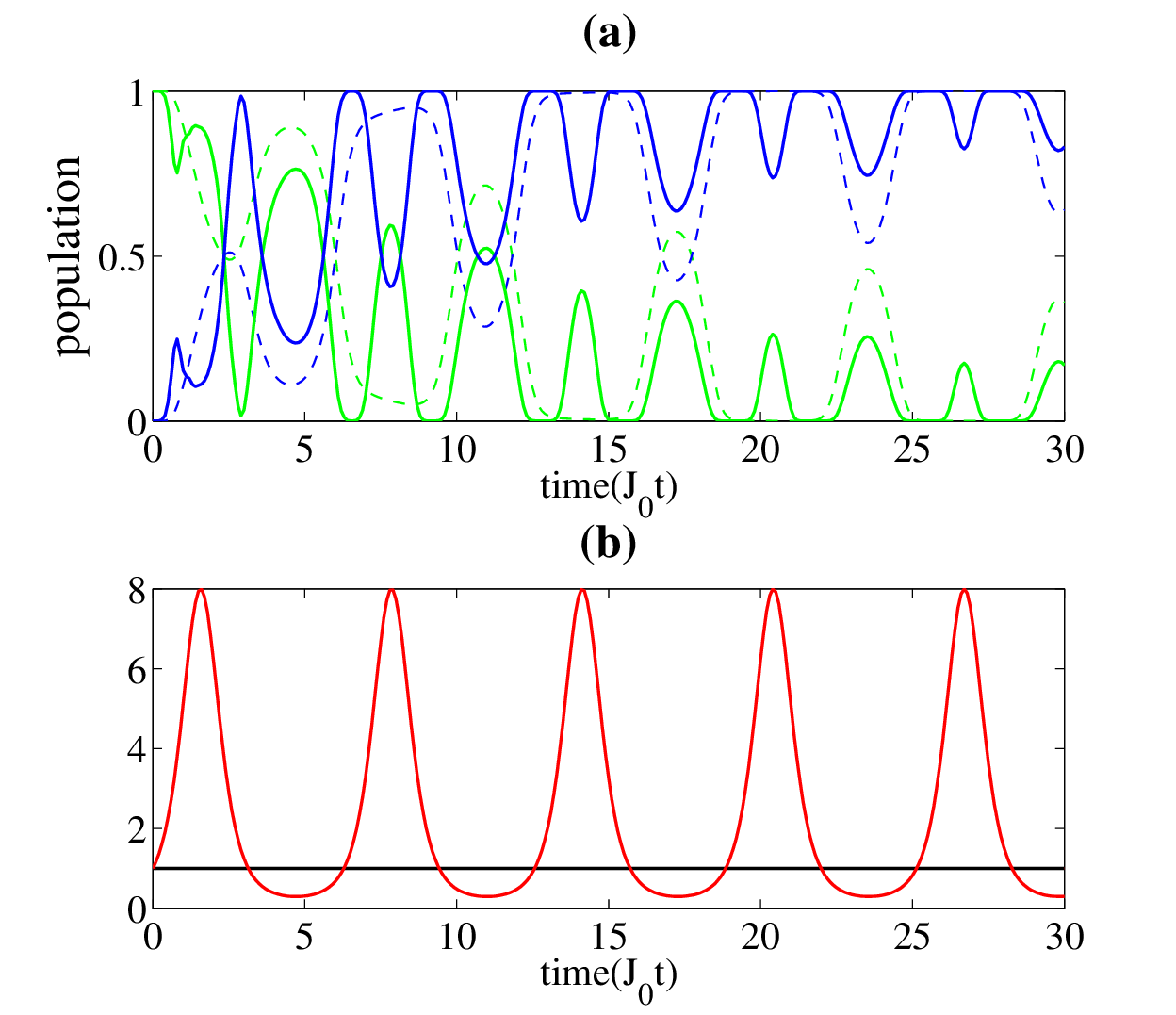}
\caption{} \label{Fig2}
\end{figure}

\newpage
Fig. 6. (a) Populations of the sinks for the optimal coherent (solid blue curve) and optimal incoherent (dashed blue curve) transport, when the sites 1 and 4 oscillate around their equilibrium positions with the phase difference $\Delta \phi=\pi$, along the horizontal line with assumptions of $a=1/4$ and $\omega_{0}=1$ for each of the oscillations. (b) Time dependence of the coupling strength $\zeta_{1}(t):=J_{1,2}(t)=J_{1,3}(t)$ (solid red curve) and $\zeta_{2}(t):=J_{2,4}(t)=\pm J_{3,4}(t)$ (dotted-dashed black curve).
\begin{figure}
\centering
\includegraphics[width=445 pt]{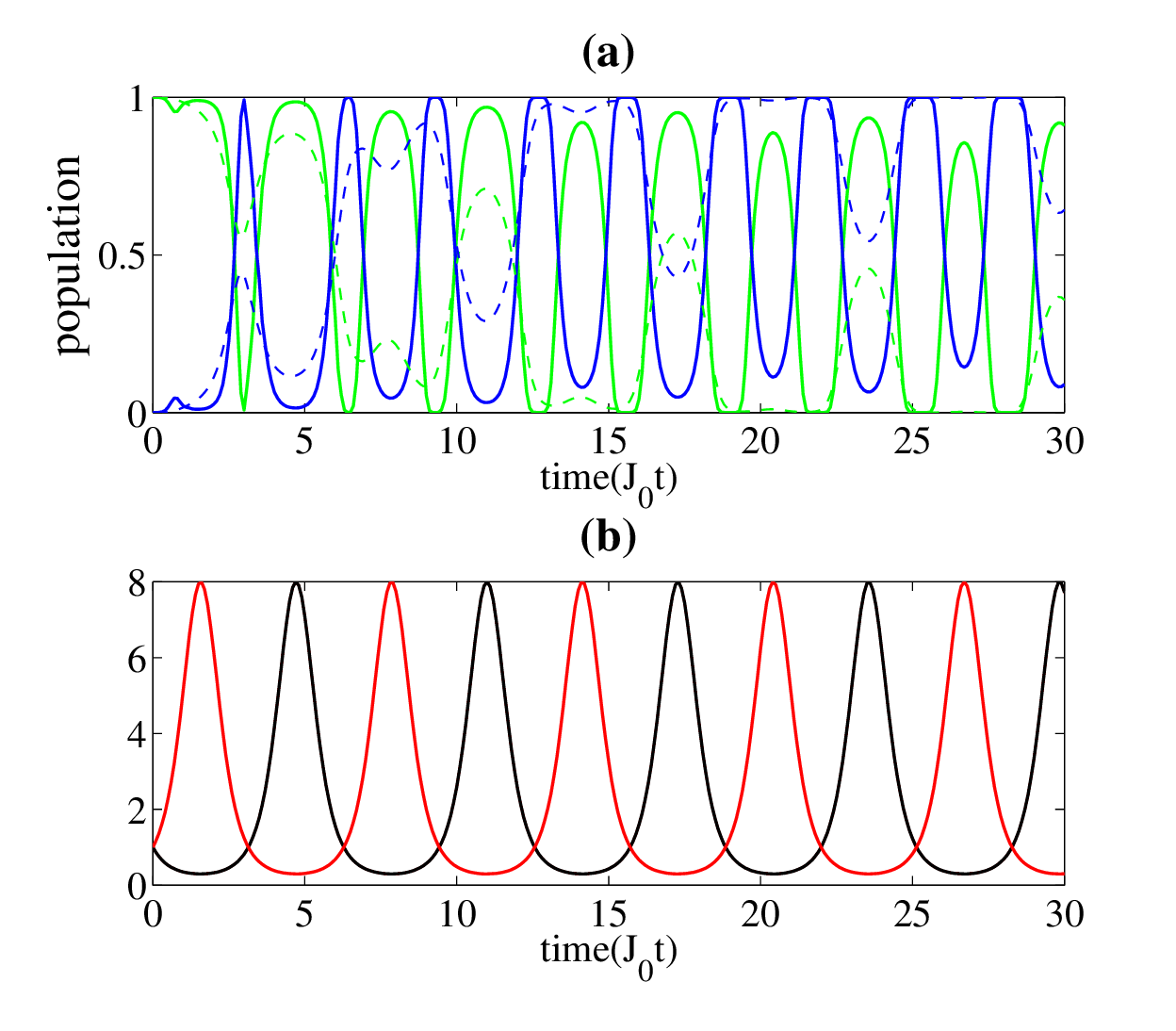}
\caption{} \label{Fig3}
\end{figure}

\newpage
Fig. 7. (a) Populations of the sinks for the optimal coherent (solid blue curve) and optimal incoherent (dashed blue curve) transport, when the sites 2 and 3 (or site1 and 4) oscillate around their equilibrium positions with the phase difference $\Delta \phi=0$ along the vertical line (horizontal line) with assumptions of $a=1/4$ and $\omega_{0}=1$ for each of the oscillations. (b) Time dependence of the coupling strength $\zeta_{1}(t):=J_{1,2}(t)=J_{1,3}(t)$ and $\zeta_{2}(t):=J_{2,4}(t)=\pm J_{3,4}(t)$ with $\zeta(t):=\zeta_{1}(t)=\zeta_{2}(t)$.
\begin{figure}
\centering
\includegraphics[width=445 pt]{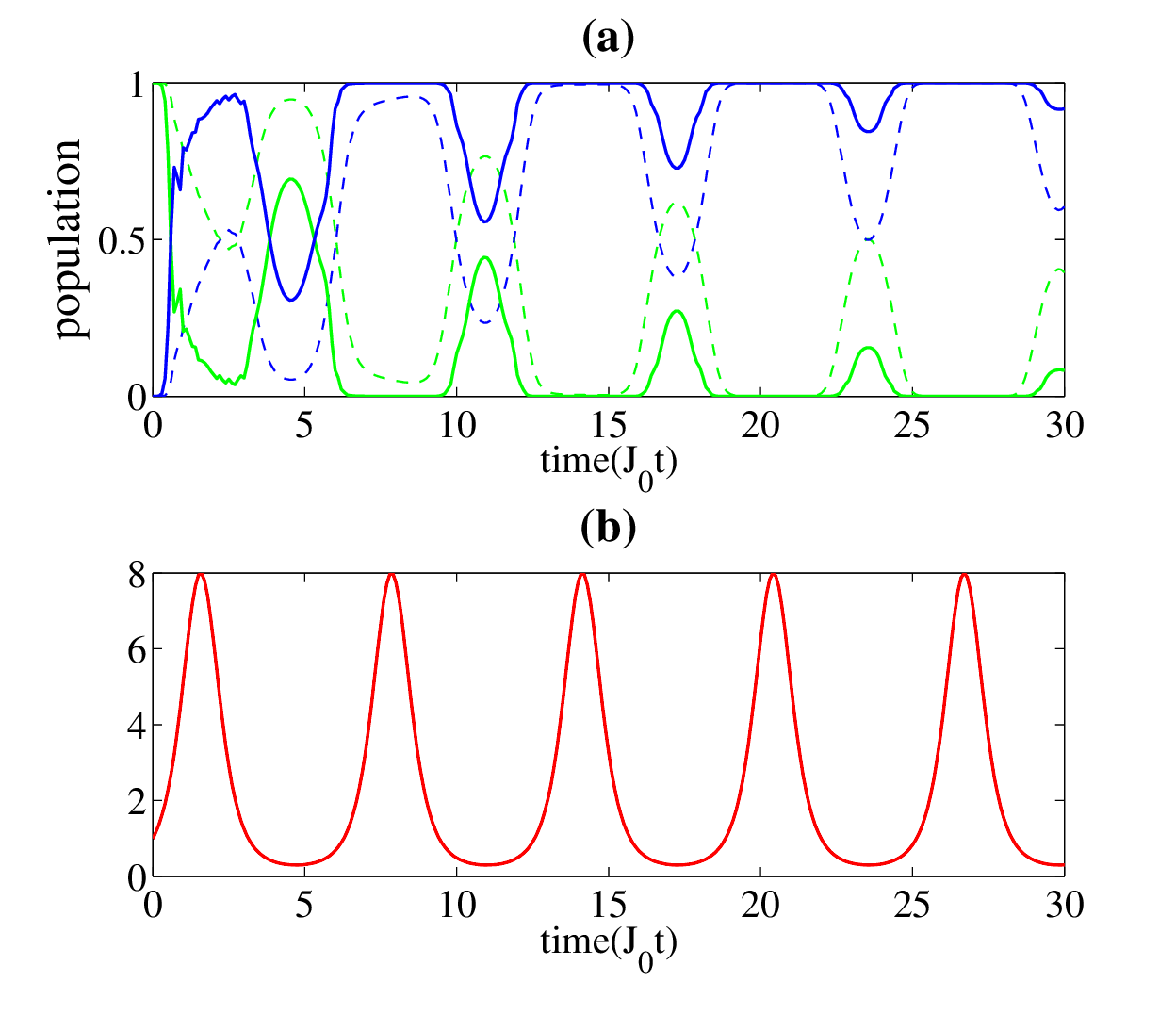}
\caption{} \label{Fig1}
\end{figure}

\end{document}